\documentclass{llncs}
\usepackage{wrapfig}
\usepackage{etex}
\usepackage[section]{placeins}
\usepackage[english]{babel} 
\usepackage[latin1]{inputenc} 
\usepackage{latexsym}
\usepackage{epsfig} 
\usepackage{listings}
\usepackage{url,xspace} 
\usepackage[draft]{fixme}
\usepackage{color} 
\usepackage{multirow}
\usepackage{latexsym,amssymb}
\usepackage{fancybox,epic}
\usepackage[noend]{algpseudocode}
\usepackage{amsmath}
\usepackage{graphics}
\usepackage{graphicx}
\usepackage{todonotes}
\usepackage{hyperref}

\usepackage{graphicx}
\usepackage{xcolor}
\definecolor[named]{ACMBlue}{cmyk}{1,0.1,0,0.1}
\definecolor[named]{ACMYellow}{cmyk}{0,0.16,1,0}
\definecolor[named]{ACMOrange}{cmyk}{0,0.42,1,0.01}
\definecolor[named]{ACMRed}{cmyk}{0,0.90,0.86,0}
\definecolor[named]{ACMLightBlue}{cmyk}{0.49,0.01,0,0}
\definecolor[named]{ACMGreen}{cmyk}{0.20,0,1,0.19}
\definecolor[named]{ACMPurple}{cmyk}{0.55,1,0,0.15}
\definecolor[named]{ACMDarkBlue}{cmyk}{1,0.58,0,0.21}
\hypersetup{colorlinks,
  linkcolor=ACMDarkBlue,
  citecolor=ACMPurple,
  urlcolor=ACMDarkBlue,
  filecolor=ACMDarkBlue}

\newcommand{\gasol}{\tname{Gasol}}

\definecolor{deepblue}{rgb}{0,0,0.5}
\definecolor{deepred}{rgb}{0.6,0,0}
\definecolor{deepgreen}{rgb}{0,0.5,0}
\definecolor{halfgray}{gray}{0.55}
\definecolor{ipythonframe}{RGB}{207, 207, 207}

\definecolor{ckeyword}{HTML}{7F0055}
\definecolor{ccomment}{HTML}{3F7F5F}
\definecolor{cnumber}{HTML}{2A0099}
\definecolor{pblue}{rgb}{0.13,0.13,1}
\definecolor{pgreen}{rgb}{0,0.5,0}
\definecolor{pred}{rgb}{0.9,0,0}
\definecolor{pgrey}{rgb}{0.46,0.45,0.48}

\definecolor[named]{ACMBlue}{cmyk}{1,0.1,0,0.1}
\definecolor[named]{ACMYellow}{cmyk}{0,0.16,1,0}
\definecolor[named]{ACMOrange}{cmyk}{0,0.42,1,0.01}
\definecolor[named]{ACMRed}{cmyk}{0,0.90,0.86,0}
\definecolor[named]{ACMLightBlue}{cmyk}{0.49,0.01,0,0}
\definecolor[named]{ACMGreen}{cmyk}{0.20,0,1,0.19}
\definecolor[named]{ACMPurple}{cmyk}{0.55,1,0,0.15}
\definecolor[named]{ACMDarkBlue}{cmyk}{1,0.58,0,0.21}

\lstdefinelanguage{Solidity}{ keywords=[1]{anonymous, synchronized,
	assembly, assert, break, case, catch, class, constant,
	continue, contract, debugger, default, delete, do, else,
	event, export, external, finally, for, function, if,
	implements, import, in, indexed, instanceof, interface,
	internal, is, length, library, memory, modifier, new, payable,
	pragma, private, protected, public, pure, require, returns,
	send, storage, struct, super, switch, then, throw, transfer,
	try, typeof, using, view, while, with, ecrecover}, 
	keywordstyle=[1]\color{blue}\bfseries\scriptsize,
	keywords=[2]{bool, bytes, bytes1, bytes2, bytes3, bytes4, bytes5, bytes6, bytes7, bytes8, bytes9, bytes10, bytes11, bytes12, bytes13, bytes14, bytes15, bytes16, bytes17, bytes18, bytes19, bytes20, bytes21, bytes22, bytes23, bytes24, bytes25, bytes26, bytes27, bytes28, bytes29, bytes30, bytes31, bytes32, enum, int, int8, int16, int24, int32, int40, int48, int56, int64, int72, int80, int88, int96, int104, int112, int120, int128, int136, int144, int152, int160, int168, int176, int184, int192, int200, int208, int216, int224, int232, int240, int248, int256, mapping, string, uint, uint8, uint16, uint24, uint32, uint40, uint48, uint56, uint64, uint72, uint80, uint88, uint96, uint104, uint112, uint120, uint128, uint136, uint144, uint152, uint160, uint168, uint176, uint184, uint192, uint200, uint208, uint216, uint224, uint232, uint240, uint248, uint256, var, void, ether, finney, szabo, wei, days, hours, minutes, seconds, weeks, years},	
	keywordstyle=[2]\color{teal}\bfseries\scriptsize,
	keywords=[3]{block,
        this, true, false, msg, data, sender, value, sig,
	value, now, tx},	
    keywordstyle=[3]\color{violet}\bfseries\scriptsize,
    identifierstyle=\color{black}, sensitive=false, comment=[l]{//},
    morecomment=[s]{/*}{*/}, commentstyle=\color{gray}\ttfamily,
    stringstyle=\color{red}\ttfamily, morestring=[b]', morestring=[b]",
    	basicstyle=\scriptsize\sffamily,                   
}

\newcommand{\lst}[1]{\lstinline!#1!}

\definecolor{shadecolor}{gray}{1.00}
\definecolor{ddarkgray}{gray}{0.75}
\definecolor{darkgray}{gray}{0.30}
\definecolor{light-gray}{gray}{0.87}

\newcommand{\etal}{\emph{et~al.}\xspace}

\newcommand{\tname}[1]{\textsc{#1}\xspace}

\newcommand{\plname}[1]{\textsf{#1}\xspace}

\newcommand{\solidity}{\plname{Solidity}}

\newcommand{\secbeg}{\vspace*{-0.15cm}}

\makeatletter
\newsavebox\myboxA
\newsavebox\myboxB
\newlength\mylenA

\newcommand*\xoverline[2][0.75]{%
    \sbox{\myboxA}{$\m@th#2$}%
    \setbox\myboxB\null
    \ht\myboxB=\ht\myboxA%
    \dp\myboxB=\dp\myboxA%
    \wd\myboxB=#1\wd\myboxA
    \sbox\myboxB{$\m@th\overline{\copy\myboxB}$}
    \setlength\mylenA{\the\wd\myboxA}
    \addtolength\mylenA{-\the\wd\myboxB}%
    \ifdim\wd\myboxB<\wd\myboxA%
       \rlap{\hskip 0.7\mylenA\usebox\myboxB}{\usebox\myboxA}%
    \else
        \hskip -0.5\mylenA\rlap{\usebox\myboxA}{\hskip 0.7\mylenA\usebox\myboxB}%
    \fi}
\makeatother

\lstdefinestyle{numbers}
{numbers=left, numberstyle=\tiny}

\usepackage{epsfig}

\usepackage{booktabs}
 \usepackage[all]{xy}

\usepackage{amssymb}
\usepackage{latexsym}
\usepackage{fancybox,epic}
\usepackage{algorithm}
\usepackage{amsmath}
\usepackage{colortbl}

\usepackage{pgf}
\usepackage{tikz}
\usepackage{amsmath}

\let\subparagraph\paragraph

\usepackage{listings}

\title{\textsc{GASOL}: Gas Analysis and Optimization for Ethereum Smart Contracts}
 \author{Elvira Albert$^1$ \and Jes\'us Correas$^1$  \and  Pablo
   Gordillo$^1$ \and\\ Guillermo Rom\'an-D\'iez$^2$  \and  Albert Rubio$^1$ }

  \institute{
  Complutense University of Madrid,  Spain \and
  Universidad Polit\'ecnica de Madrid, Spain}

\begin{document}
\maketitle

\begin{abstract}
  We present the main concepts, components, and usage of \gasol, a Gas
  AnalysiS and Optimization tooL for Ethereum smart contracts. \gasol
  offers a wide variety of \emph{cost models} that allow inferring the
  gas consumption associated to selected types of EVM instructions
  and/or inferring the number of times that such types of bytecode
  instructions are executed. Among others, we have cost models to
  measure only storage opcodes, to measure a selected family of
  gas-consumption opcodes following the Ethereum's classification, to
  estimate the cost of a selected program line, etc. After choosing
  the desired cost model and the function of interest, \gasol returns
  to the user an upper bound of the cost for this function.  As the
  gas consumption is often dominated by the instructions that access
  the storage, \gasol uses the gas analysis to detect under-optimized
  storage patterns, and includes an (optional) automatic optimization
  of the selected function. Our tool can be used within an Eclipse
  plugin for \solidity which displays the gas and instructions bounds
  and, when applicable, the gas-optimized \solidity function.
\end{abstract}


\secbeg
\secbeg
\secbeg
\section{Introduction and Main Applications}\label{sec:introduction}
\secbeg

Ethereum \cite{yellow} is a global, open-source platform for
decentralized applications that has become the world's leading
programmable block\-chain.  As other block\-chains, Ethereum has a
native cryptocurrency named \emph{Ether}.  Unlike other
block\-chains, Ethereum is programmable using a Turing complete
language, i.e., developers can code smart contracts that control
digital value, run exactly as programmed, and are immutable.  A smart
contract is basically a collection of code (its functions) and data
(its state) that resides at a specific address on the Ethereum
blockchain.  Smart contracts on the Ethereum blockchain are metered
using \emph{gas}.  Gas is a unit that measures the amount of computational
effort that it will take to execute each operation. Every single
operation in Ethereum, be it a transaction or a smart
contract instruction execution, requires some amount of gas.  The gas
consumption of the Ethereum Virtual Machine (EVM) instructions is
spelled out in \cite{yellow}; importantly, instructions that use
replicated storage are gas-expensive.  Miners get paid an amount in
\emph{Ether} which is equivalent to the total amount of gas it took
them to execute a complete operation. The rationale for gas metering
is threefold:
(i) Paying for gas at the moment of proposing the transaction
prevents the emitter from wasting miners computational power
by requiring them to perform  worthless intensive work.
(ii) Gas fees disincentive users to consume too much of replicated
\emph{storage}, which is a valuable resource in a blockchain-based
consensus system (this is why storage bytecodes are gas-expensive).
(iii) It puts a cap on the number of computations that a transaction
can execute, hence prevents DoS attacks based on non-terminating
executions.

\solidity \cite{solidity} is the most popular language to write
Ethereum smart contracts that are then compiled into EVM bytecode. The
\solidity compiler, \textbf{solc}, is able to generate only
\emph{constant} gas bounds. However, when the bounds are
\emph{parametric} expressions that depend on the function parameters, on
the contract state, or on the blockchain state (according to the
experiments in \cite{vecos19} this happens in almost 10\% of the
functions), named \textbf{solc}, returns $\infty$ as gas bound.  This paper
presents \gasol, a resource analysis and optimization tool that is
able to infer parametric bounds and optimize the gas consumption of
Ethereum smart contracts.  \gasol takes as input a smart contract
(either in EVM, disassembled EVM, or in \textsf{Solidity} source
code), a selection of a cost model among those available in the system
(c.f. Section~\ref{sec:gas-analysis-using}), and a selected
public function, and it automatically infers \emph{cost upper bounds}
for this function.  Optionally, the user can enable the gas
optimization option (c.f. Section~\ref{sec:appl-gas-optim}) to
optimize the function w.r.t.\ storage usage, a highly valuable
resource. \gasol has a wide range of applications: (1) It can be used
to estimate the gas fee for running transactions, as it soundly
over-approximates the gas consumption of functions. (2) It can be used
to certify that the contract is free of out-of-gas vulnerabilities, as
our bounds ensure that if the gas limit paid by the user is higher
than our inferred gas bounds, the contract will not run
out-of-gas. (3) As an attacker, one might estimate, how much
\emph{Ether} (in gas), an adversary has to pour into a contract in
order to execute an out-of-gas attack. Also, attacks were produced by
introducing a very large number of underpriced bytecode instructions
\cite{DBLP:journals/corr/abs-1909-07220}. Our cost models could allow
detecting these second type of attacks by measuring how many
instructions will be executed (that should be very large) while its
associated gas consumption remains very low. (4) As we will show in
the paper,
the gas analysis can be used to detect gas-expensive fragments of code
and automatically optimize them.



\secbeg
\secbeg

\section{Gas Analysis using \gasol}\label{sec:gas-analysis-using}
\secbeg

\begin{figure}[t]
\secbeg
\secbeg
\secbeg
\secbeg
\begin{center}
\fbox{\includegraphics[scale=0.3]{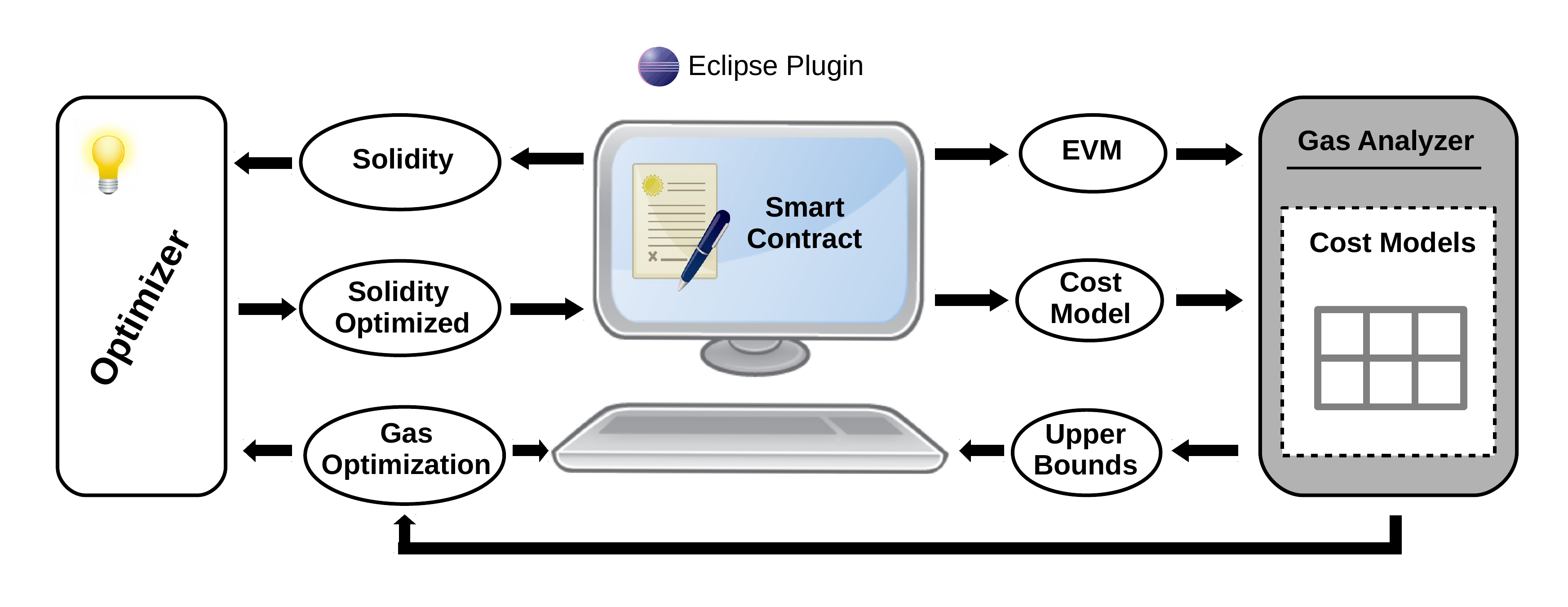}}
\secbeg \secbeg \caption{Overview of \gasol's components}\label{fig:architecture}
\end{center} 
\secbeg \secbeg \secbeg \secbeg\end{figure}

Figure~\ref{fig:architecture} overviews the components of the \gasol
tool. The programmer can use \gasol during the software development
process from its Eclipse plugin that allows selecting the cost model
of interest and the function to be analyzed and/or optimized from the
Outline. This selection together with the compiled EVM code is sent to
the gas analyzer. A technical description of all phases that comprise
a gas analysis for EVM smart contracts is given in
\cite{vecos19}. Basically, the analyzer uses various tools
\cite{Oyente,AlbertGLRS18} to extract the CFGs and decompile them into
a high-level representation from which upper bounds (UB) are produced
by using extensions of resource analyzers and solvers
\cite{AlbertAFGGMPR14,pubs}. However, in our basic gas analyzer named \tname{gastap} \cite{vecos19}, there was
only one cost model to compute the overall gas consumption of the
function (including the opcode and memory gas costs \cite{yellow}),
while \gasol is an extension of \tname{gastap} that 
introduces optimization,  a wide variety of analysis options to define novel cost
models, and an Eclipse plugin.
The UBs are provided to the user in the console as well as in markers
for functions within the Eclipse editor. If the user had selected the
optimization option, the analyzer detects potential sources of
optimization and feeds them to the optimizer to generate an optimized
\solidity function within a new file. 


Fig.~\ref{fig:example} displays our Eclipse plugin that contains a
fragment of the public smart contract
\textsf{ExtraBalToken}~\cite{ExtraBalToken} 
used as
running example. We can see its six state variables and its function
\textsf{fill} that we will analyze and optimize. The right side window
shows \gasol's configuration options to set up the \emph{cost model}:

\noindent \emph{(i) Type of resource (gas/instructions)}: by selecting
\emph{gas}, we estimate the gas consumption according to the gas model
in \cite{yellow} (hence, use \gasol as a gas analyzer); by selecting
\emph{instructions}, we estimate the number of bytecode instructions
executed (using \gasol as a standard complexity analyzer).

\noindent \emph{(ii) Type of instructions}: allows selecting which instructions (or
  group of instructions) will be measured as follows.

\begin{itemize} \secbeg\secbeg
\item \emph{All}: every bytecode instruction will be measured. For
  instance, by selecting gas in (i), the function \textsf{fill}, and
  this option, we obtain as gas bound: $1077+40896 \cdot
  data$. Besides, by using this option, \gasol also yields the
  so-called memory gas (see\cite{yellow}):
  $3\cdot(data+5)+\left\lfloor{\frac{(data+5)^2}{512}}\right\rfloor$. The
  analyzer abstracts arrays by their length, hence, these bounds
  are functions  of the \emph{length of the input array}
  (denoted as $data$) and can be used, e.g., to determine precisely how much gas is
  necessary to run a transaction that executes this function.

\item \emph{Gas-family}: \cite{yellow} classifies bytecode
  instructions according to their gas consumed in six groups: zero,
  base, verylow, low, mid and high.  Instructions that do not belong
  to any of the previous groups are considered as single families.
  This option provides the cost due to each gas-family separately and,
  by using the filter in (iii), we can type the name of the desired
  group(s). As an example, for the function \textsf{fill} using gas in
  (i), we obtain gas bounds $297+315\cdot data$ and $16+8 \cdot data$
  for the gas-families verylow and mid, resp.
 
\item \emph{Storage}: only the instructions that access the storage
  (namely bytecodes \texttt{SLOAD} and \texttt{SSTORE}) are
  accounted. The gas bounds displayed within the Eclipse console in
  Fig.~\ref{fig:example} correspond to this setting, where we can see
  that the gas due to the access of each basic storage variable is shown
  separately. The first row \textsf{unknown} accumulates the gas
  of all
  accesses to non-basic types (data structures) as we still cannot
  identify them.   By comparing this storage gas with the overall gas
  bound shown above for \emph{All}, we can observe that most of the
  gas consumed by the function is indeed dominated by the storage
  (namely 40.000 out of 40.896 at each loop iteration) and
  it is thus a target for optimization, as we will see in
  Sec.~\ref{sec:appl-gas-optim}.

\item \emph{Storage-optimization}: it bounds the number of
  \texttt{SLOAD} and \texttt{SSTORE} instructions executed by the
  current function (excluding those in transitive calls). It is the
  cost model that is used to detect and carry out the optimization
  described in Sec.~\ref{sec:appl-gas-optim}. Thus, it is the only
  selection that enables the \emph{Gas optimization} that appears as
  third option, and forces the selection of ``instructions'' as type
  of resource in
  (i). We obtain for the state variable \texttt{totalSuply} the
  bound: $2\cdot data$, which captures that we execute two
  accesses (one read, one write) to field \texttt{totalSuply} at each loop iteration.

\item \emph{Line}: this option allows specifying the line number (of
  the \texttt{Solidity} program) whose cost will be measured, and the
  remaining lines will be filtered out. For instance, if the line
  number specified in the filter (iii) is 17, i.e., the \solidity instruction:
  \lst{uint amount  = data[i] / D160}, the obtained gas bound is 
  $3+97\cdot data$. In the absence of number in the filter, the
  bounds are given separately for all program lines. This option is
  intended to help the programmer in improving the gas consumption of
  her code by trying out different implementation options and
  comparing the results.
\begin{figure}[t]
\begin{center}
  \includegraphics[scale=0.23]{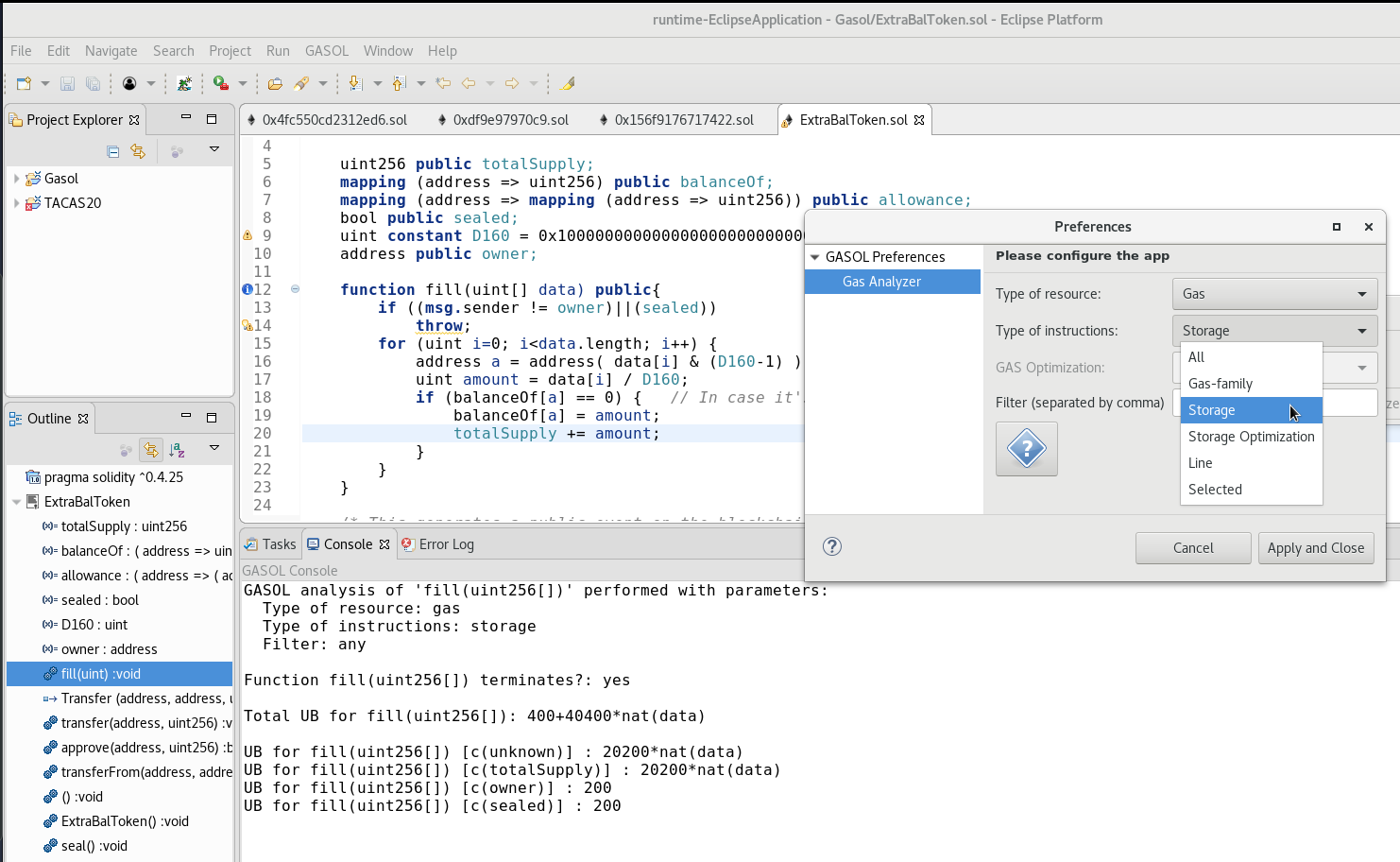}
\end{center}
\secbeg \secbeg \caption{Excerpt of smart contract
  \texttt{ExtraBalToken} in \solidity within Eclipse plugin. View of
  analyze+configure menu for selection of cost model and optimization
  flag.}\label{fig:example} \secbeg \secbeg \secbeg
\secbeg \end{figure}

\item \emph{Selected}: allows computing the consumption associated to
  each different EVM instruction  separately. For instance,
  if we select the bytecode instructions \texttt{MLOAD} and
  \texttt{SHA3}, we obtain the gas bounds $6+15 \cdot data$ and
  $84 \cdot data$ resp. As in the previous option, the filter allows
  the user to select the bytecode instructions of interest and filter
  out the remaining.
\end{itemize}
\noindent \emph{(iii) Filter}: this is a text field used to filter out
  information from the UBs.  For \emph{gas-family}, the user
  can specify low, mid, etc. For \emph{storage}, it allows specifying the
  name of the basic field(s) whose storage will be measured. For \emph{line}
  and \emph{selected}, we can type the line numbers and names of
  bytecode instructions of interest.
%
 Once all options have been selected, we have set
up a cost model that is sent together with the EVM code to the gas
analyzer and, after analysis, it outputs an UB for the
selected function w.r.t.\ the cost model activated by the
options. This UB is displayed, as shown in Fig.~\ref{fig:example} in
the console of the Eclipse plugin, and also within markers next to
the function definition.


\secbeg
\secbeg
\section{Gas Optimization using \gasol}\label{sec:appl-gas-optim}
\secbeg
\secbeg

The information yield by the gas analysis is used in \gasol to detect
potential optimizations. Currently, the optimization target is the
reduction of the gas consumption associated to the usage of
storage. In particular, we aim at replacing multiple accesses to the
same (global) storage data within a fragment of code (each write
access costs 20.000 in the worst case and 5.000 in the best case) by one access that copies the
data in storage to a (local) memory position followed by accesses to
such memory position (an access to the local memory costs only 3) and
a final update to the storage if needed. The cost model number of
instructions for storage-optimization described in
Sec.~\ref{sec:gas-analysis-using} allows us to detect such storage
optimizations, namely for each different field, if we get a bound that
is different from one, we know that there may be multiple accesses to the
same position in the storage and we try to replace them by
gas-efficient memory accesses.  Our transformation is done at the
level of the \solidity code, by defining a local variable with the same
name as the state variable to transform, and introducing setter and
getter functions to access the storage variable. Currently, we can
transform accesses to variables of basic types, in the future, we plan
to extend it to data structures (maps and arrays). The
number of instructions bound  for field
\texttt{totalSupply} is $2 \cdot data$ (hence $\neq 1$), and our optimization of \texttt{fill} is:
\begin{center}
\begin{tabular}{ll}
\begin{lstlisting}[name=code,language=Solidity]
 function fill(uint[] data) {
   uint256 totalSupply = get_field_totalSupply(); 
     
   if ((msg.sender != owner)||(sealed))
      throw;
   for (uint i=0; i<data.length; i++) {
     address a = address( data[i] & (D160-1) );
\end{lstlisting}
&
\hspace{0.6cm}
\begin{lstlisting}[name=code,language=Solidity]
     uint amount = data[i] / D160;
     if (balanceOf[a] == 0) {  
       balanceOf[a] = amount;
       totalSupply += amount;
     }
   } 
  set_field_totalSupply(totalSupply);   
 }
\end{lstlisting}
\end{tabular}
\end{center}
\medskip The gas bound (using the option \emph{All}) for the optimized
\textsf{fill} yield by \gasol is $21368+20674\cdot data$, which means
that, assuming the worst case for
write access to storage, the gas consumed inside the loop is 49.45\% smaller than the one for the original
\textsf{fill} function (the memory gas does not change). Note that,
even if we consider the best case of 5.000 for write access to storage
for the accesses we have optimized, the gas reduction is still around
 20\%. This is, in fact, what we have manually estimated using the
actual data of the 82 times this function has been executed in the
Ethereum blockchain, achieving with \gasol a total saving of almost 60M gas. As our transformation is local to the function, in order to
be sound, we check that the transformed global data is not being accessed
by transitive calls. For instance, if there was a call to another function
from function \textsf{fill} that accesses \texttt{totalSupply}, we would not
transform it. Besides, for efficiency, we check if all accesses are read
(bytecode \texttt{SLOAD}) and, in such case, we do not need to invoke the setter
at the end (and avoid an unnecessary write access).


\secbeg
\secbeg
\secbeg
\section{Related Tools and Conclusions}
\secbeg
\secbeg
\secbeg

Numerous tools are being developed to catch different types of
vulnerabilities of smart
contracts~\cite{Luu-al:CCS16,GrossmanAGMRSZ18,Nikolic-al:Maian,KruppR18,Kalra-al:NDSS18,TsankovDDGBV18,Kolluri-al:laws,Bhargavan-al:PLAS16,Grishchenko-al:POST18,Amani-al:CPP18}.
As mentioned in Sec.~\ref{sec:introduction}, the \solidity compiler
\texttt {solc} is not able to give any gas estimation for the running
example, as its gas consumption is not constant. Therefore, new gas
analysis tools are being developed to detect potential gas related
vulnerabilities and to infer bounds in these complex situations. The
purpose of the \tname{Gasper} and \tname{MadMax} tools is precisely
the detection of gas related vulnerabilities.  \tname{MadMax}
\cite{madmax} focuses on identifying control- and data-flow patterns
inherent for the gas-related vulnerabilities, thus, it works as a
bug-finder, rather than as a gas analyzer like \gasol.  Similarly,
\tname{Gasper}  identifies gas-costly programming
patterns~\cite{ChenLLZ17} by matching specific control-flow patterns
and using SMT
 solvers and symbolic computation.
Thus, it is an optimization
detector, not an automatic optimizer as \gasol.  The recently
developed \texttt{ebso} tool \cite{superoptimizer} also aims at
optimizing the gas consumption of EVM code. In contrast to \gasol,
\texttt{ebso}'s optimizations are limited to a basic block level,
while our transformation might involve several blocks of the CFG and
would not be achievable by \texttt{ebso}'s approach. Also,
\texttt{ebso} is not guided by the results of an automatic resource
analysis which can capture the expensive storage patterns as in our
case. Instead it is based on a full exploration of all possible
alternative instructions (within the considered block) that would lead
to the same result and consume less gas. They have
obtained a number of rewrite rules that define sequences of bytecode
instructions that can be replaced by equivalent ones that consume
less.  We could easily incorporate such basic block replacement
optimizations within our tool, and it is part of our agenda.
 
The approach of \cite{Marescotti-al:ISoLA18}, like ours, aims at
inferring precise gas bounds.
Their approach is based on symbolically enumerating all execution
paths~\cite{Biere-al:TACAS99} and unwinding loops to a limit.
Instead, using resource analysis, \gasol infers the maximal number of
iterations for loops and generates accurate gas bounds which are valid
for any possible execution of the function and not only for the
unwound paths.  The approach by Marescotti~\etal has not been
implemented in the context of EVM and a tool like \gasol has not been
delivered.  An orthogonal line of work with ours is the construction
of resource-oriented attacks \cite{DBLP:journals/corr/abs-1909-07220}
that exploit the weaknesses of the EVM gas model. \gasol's cost models
could help detect this resource-oriented attacks by estimating the number of
executed bytecode instructions (very high) and their associated gas
consumption (very low).

Finally, there is a tendency to define new languages (see Scilla
\cite{oopsla19}, Michelson \cite{tezos}) for programming smart
contracts that provide certain safety guarantees, e.g., Scilla
\cite{oopsla19} provides predictable gas consumption by disallowing
general recursion and while-loops.  However, Ethereum is today the
most widely used blockchain, and \solidity the most popular
programming language to write Ethereum smart contracts, for which a
gas analyzer+optimizer is of clear relevance.

\bibliographystyle{plain}
\bibliography{biblio}



\end{document}